\documentclass[lnbip]{svmultln}

\newcommand{\rb}[1]{\rotatebox{90}{\textit{#1}}}

\usepackage{color}
\usepackage{graphics}
\usepackage{multirow}
\usepackage{array,ragged2e}

\newcolumntype{C}[1]{>{\Centering\arraybackslash\hspace{0pt}}m{#1}}
\newcolumntype{R}[1]{>{\RaggedLeft\arraybackslash\hspace{0pt}}m{#1}}
\newcolumntype{L}[1]{>{\RaggedRight\arraybackslash\hspace{0pt}}m{#1}}

\newtheorem{rquestion}{RQ}

\begin{document}

\mainmatter              
\title{On the Benefit of Automated Static Analysis for Small and Medium-Sized Software Enterprises}

\titlerunning{On the Benefit of Automated Static Analysis for SMEs}  
%
\author{Mario Gleirscher\inst{1} \and Dmitriy Golubitskiy\inst{1} \and Maximilian Irlbeck\inst{1} \and \\ Stefan Wagner\inst{2}}
\authorrunning{M.~Gleirscher, D.~Golubitskiy, M.~Irlbeck, S.~Wagner}   
%
\tocauthor{Mario Gleirscher, Dmitriy Golubitskiy, Maximilian Irlbeck, Stefan Wagner}
\institute{Institut f\"ur Informatik, Technische Universit\"at M\"unchen, 
Germany\\
\email{{gleirsch,golubits,irlbeck}@in.tum.de}
\and
Software Engineering Group, Institute of Software Technology,\\ University of Stuttgart, Germany\\
\email{stefan.wagner@informatik.uni-stuttgart.de}
}

\maketitle              

\index{Gleirscher, Mario} 
\index{Wagner, Stefan}  
\index{Irlbeck, Maximilian}
\index{Golubitskiy, Dmitriy}

\begin{abstract}        
Today's small and medium-sized enterprises (SMEs) in the software industry are faced with major challenges. While having to work efficiently using limited resources they have to perform quality assurance on their code to avoid the risk of further effort for bug fixes or compensations. Automated static analysis can reduce this risk because it promises little effort for running an analysis. We report on our experience in analysing five projects from and with SMEs by three different static analysis techniques: code clone detection, bug pattern detection and architecture conformance analysis. We found that the effort that was needed to introduce those techniques was small (mostly below one person-hour), that we can detect diverse defects in production code and that the participating companies perceived the usefulness of the presented techniques as well as our analysis results high enough to include the techniques in their quality assurance.

\end{abstract}

\vspace{-7pt}
\section{Introduction}
\label{sec:intro}

Small and medium-sized enterprises (SMEs) play a decisive role in global software industry. In many countries, like the US, Brazil or China, these companies represent up to 85\% of all software organisations \cite{Richardson2007} and carry out the majority of software development \cite{MISHRA}. Nevertheless, SMEs are confronted with special circumstances like limited resources, lack of expertise or financial insecurity. 

\paragraph{Problem}
While there are many articles focusing on process improvement in SMEs~\cite{Kautz1999,MISHRA,Wangenheim2006}, we found no study that looks at specific quality assurance (QA) techniques and their application in this context. Contrary to this observation, the properties of automated static analysis techniques seem to be suitable for SMEs. The benefits of such techniques lie in their low-cost application and their potential to detect critical quality defects \cite{Zheng06,ayewah07}. Such defects are risky for the further development and increase costs. These arguments are promising for small software enterprises and their need for efficient quality assurance.

\paragraph{Research Objective} Our goal is to answer the question whether SMEs can benefit from automated static analysis techniques. Is it possible to introduce a set of such techniques in their existing projects with low effort? What kind of defects can be found using these techniques? Finally, is the perceived usefulness for the enterprises high enough to justify the needed effort? We think that these aspects are useful for future decisions in SMEs on using static analysis techniques in their projects.

\paragraph{Contribution}
In this article, we describe our experience in analysing five projects of five SMEs using three different static analysis techniques: code clone detection, bug pattern detection and architecture conformance analysis. We evaluate the effort that is needed to introduce these techniques, the pitfalls we came across and how the participating enterprises evaluated the presented techniques as well as the defects we discovered in their projects.

\vspace{-7pt}
\section{Approach}
\label{sec:approach}

We describe our experiences with transferring static analysis technology to
small and medium-sized enterprises. This section illustrates the research context, 
i.e., the participating enterprises, our guiding research questions, the 
regarded static analysis techniques, the procedure we used to get answers 
to the research questions and finally the study objects we employed to gather 
the experiences.

\subsection{Research Context}
\label{sec:context}

Fundamental for our research was the collaboration with five SMEs, 
all resident in the Munich area and selected through personal contacts and a 
series of information events and workshops. Details regarding the selection 
process can be found in Sec.~\ref{sec:proc}.
Following the definition of the European Commission~\cite{EuComm2003}, one 
of the participating enterprises is micro-, two are small and 
two are medium-sized considering their staff head count and annual turnover.
The presented research is based on the experience with these enterprises 
gathered in a project from March 2010 to April 2011.

\subsection{Research Questions}
\label{sec:questions}

Our overall research objective is to analyse the transfer of new 
and innovative quality assurance techniques to small enterprises. 
We structure this objective into two major research questions.

\begin{rquestion}
What problems occur while introducing and applying static analysis techniques at SMEs?
\end{rquestion}

SMEs exhibit special characteristics, such as 
generalist employees instead of specialists for quality assurance. Hence,
smooth introduction and application are necessary so that the enterprises can
adopt and make use of static analysis. We further break this down into
two sub-questions:

\noindent
\textbf{RQ 1.1} \emph{What technical problems occur?}

Static analysis is tightly coupled to tools that perform and report the
analysis. Hence, the ease to introduce and apply static analysis also
depends on how many and which technical problems the software engineers
need to solve.

\noindent
\textbf{RQ 1.2} \emph{How much effort is necessary?}

If the effort necessary to bring the analyses up and running is too large, 
it can be a killer criterion for an SME, which cannot afford 
to reserve extra capacities for that. Therefore, we analyse the effort spent 
in the introduction and application.

\begin{rquestion}
How useful are static analysis techniques for SMEs?
\end{rquestion}

Beyond how easy or problematic it is to introduce and apply static
analysis in SMEs, we are interested in whether we can
produce useful results for them. Even a small effort should not be
spent if there is no return on investment. We again break this
question down into two sub-questions:

\noindent
\textbf{RQ 2.1} \emph{Which defects can be found?}

We establish a measure of usefulness by analysing
the types and numbers of defects found by using the static analysis
tools at the SME. If critical defects can be found, the application 
of the techniques is considered useful. We neither focus on specification defects and whether they can be found at all, nor do we perform cause and effects analyses for defects except for some criticality assessments.

\noindent
\textbf{RQ 2.2} \emph{How do the companies perceive the usefulness?}

We add the subjective perception of our project partners.
How do they interpret the results of the static analysis tools? Do they believe they can work
with those tools and are they going to apply them continuously in their future projects? This way, we augment the information we gained from defect analysis.

\renewcommand\theenumi{\arabic{enumi}}
\renewcommand\labelenumi{\theenumi)}

\subsection{Static Analysis Techniques}
\label{sec:methods}

Static analysis is known as the checking of software against certain properties without executing it. It includes manual techniques, such as
reviews and inspections, as well as automated techniques.
As manual analyses are time-consuming and prone to missing problems in the
huge amount of code to analyse, automation has high potential. For example,
to detect simple and reoccurring problems in source code, such as using ``==''
instead of ``equals'' to compare strings in Java, should not be the task
of human reviewers. They should concentrate on the more subtle and domain-related problems. From the interviews with our partners
and the experiences at our research groups, we chose
three important techniques, which we introduce in detail
in the following. Technically, we employ the open-source tool ConQAT\footnote{\url{http://www.conqat.org}} for code clone detection and architecture conformance analysis as well as for results processing of bug pattern detection.

\paragraph{Code Clone Detection}
\label{sec:methods_clones}

Modern programming languages, particularly object-oriented ones, offer various abstraction mechanisms to facilitate reuse of code fragments, but copy-paste is still a widely employed reuse strategy. This often leads to numerous duplicated code fragments---so called clones---in software systems. As stated in the surveys of Koschke \cite{Koschke2007} and Roy and Cordy \cite{Roy2007}, cloning is problematic for software quality for several reasons:

\begin{itemize}
\item Cloning unnecessarily increases program size and thus efforts for size-related activities like inspections and testing.
\item Changes, including bug fixes, to one clone instance often need to be made to the other instances as well, again increasing efforts.
\item Inconsistently performed changes to duplicated source code fragments can introduce bugs.
\end{itemize}

Code clone detection is an automated static analysis technique that focuses on finding duplicated code fragments. One of the most important metrics offered by this technique is \textit{unit coverage}, which is the probability that an arbitrarily chosen source statement (i.e.~a unit) is part of a clone. Another metric called \textit{blow-up} denotes the ratio of the unit count of the current software w.r.t.~the unit count of a hypothetical software without clones \cite{juergens2010achieving}. Moreover, two terms are important for clone detection: A \textit{clone class} defines a set of similar code fragments and a \textit{clone instance} is a representative of a clone class \cite{Juergens2009}.

We differentiate between conventional clone detection and gapped clone detection. During conventional clone detection, clones are considered to be syntactically similar copies; only variable, type, or function identifiers could be changed \cite{Koschke2007}. In contrast, gapped clone detection reveals clones with further modifications; statements could be changed, added, or removed \cite{Koschke2007}. While clones are an indicator of bad design, the difference between the two approaches is that only the results of gapped clone detection can reveal defects that lead to failures, which arise through unconscious, inconsistent changes in instances of a clone class.

Clone detection is supported by a number of free and commercial tools. The most popular of them are 
CCFinder\footnote{\url{http://www.ccfinder.net}}, 
ConQAT, 
CloneDR\footnote{\url{http://www.semanticdesigns.com/Products/Clone}},
and Axivion Bauhaus Suite\footnote{\url{http://www.axivion.com}}. The former two are free, while the latter two are commercial.

\paragraph{Bug Pattern Detection}
\label{sec:methods_patterns}

By this term we refer to a technique for automated detection of a variety of defects. 
Bug patterns have been thoroughly investigated, e.g.~in \cite{Zheng06}, and compared with other frequently used software quality assurance techniques such as code reviews or testing \cite{wagner:testcom05}. 
Bug patterns represent a scalable approach to efficiently reveal defects or possible causes thereof. Following Wagner et al.~\cite{DBLP:conf/icst/WagnerDAWS08} they can be cost-efficient after detecting only three field defects. Their detectors, aka \textit{rules}, aim at structural patterns recognisable from source code, executables and meta-data such as source code comments and debug symbols to gain as much knowledge as possible from a static perspective. This knowledge encompasses obvious bugs, rather complex heuristics for latent defects, e.g.~code clones (focused in Sec.~\ref{sec:methods_clones}), and less critical issues of coding style.

Because of the large bandwidth of defects, bug patterns are  categorised along a variety of tool-specific, non-standard criteria. A reason for that is that generally applicable defect classifications 
are rare, vague or difficult to use in practice \cite{DBLP:conf/issta/Wagner08}.
The tools used for this report classify their rules according to the consequences of findings such as security vulnerability, performance loss or functional incorrectness. By the term \textit{finding} we denote that a rule was applied at a specific location. Often, findings are themselves categorised by their severity and their confidence levels.

Many of the rules are realised by means of individual lexers and parsers, by using compiler infrastructures, or by more reusable means such as pattern or rule languages and machine-learning. Rules for latent defects and coding style often stem from abstract source code metrics as, e.g., realised in Ferzund, Ahsan, and Wotawa \cite{DBLP:conf/iwsm/FerzundAW08}. Among the wide variety of tools \cite{wiki:staticcodeanalysistools} available for bug pattern detection, free and more popular ones are, e.g., 
splint\footnote{\url{http://splint.org}} for C, 
cppcheck\footnote{\url{http://cppcheck.sourceforge.net}} for C++, 
FindBugs\footnote{\url{http://findbugs.sourceforge.net}} for Java as well as 
FxCop\footnote{\url{http://msdn.microsoft.com/en-us/library/bb429476\%28v=vs.80\%29.aspx}} for C\#.

\paragraph{Architecture Conformance}
\label{sec:methods_archconf}

The phenomenon of architectural erosion is a widely documented problem~\cite{Feilkas2009loss,Fiutem1998identifying,Rosik2008industrial}. 
Architectural knowledge erodes or even gets lost during the lifetime of a system. Accordingly, the documented and implemented architectures are drifting
apart from each other. This effect leads to a downward spiralling maintainability of the system. In some cases the effort needed to re-implement 
the whole system becomes lower than to maintain it. To counteract this situation different approaches are used to compare 
the system's implementation with its intended architecture.

Passos et al.~\cite{passos2010} identify three static concepts existing for architecture conformance analysis:
 \textit{Reflexion Models (RM)}, \textit{Source Code Query Languages (SCQL)} and \textit{Dependency Structure Matrices (DSM)}. 

\textit{Reflexion Models} Koschke and Simon \cite{Koschke2003} compare two models of a system to each other and check their conformance. The first model usually represents the intended architecture, the second one the implementation of the system~\cite{Knodel2007comparison}. The intended architecture consists of components and allowed relationships between components, expressed as rules. Each component itself can contain sub-components. The system's code is mapped to these components and then analysed for conformance to the given rules. This technique is used by the commercial tools SonarJ\footnote{\url{http://www.hello2morrow.com/products/sonarj}} and Structure101\footnote{\url{http://www.headwaysoftware.com}} as well as the open-source tools ConQAT and   dependometer\footnote{\url{http://source.valtech.com/display/dpm/Dependometer}}.

There are tools using \textit{SCQL} like Semmle.QL~\cite{DeMoor2007QL} or \textit{DSM} like Lattix~\cite{Sangal2005Lattix}, for 
the sake of brevity not further explained here. Both of these concepts rely strongly on the realisation of the system and cannot provide 
an architecture specification that is independent of the system's implementation~\cite{Deissenboeck2010}.

\subsection{Study Subjects and Objects}
\label{sec:partners}

\paragraph{Study Subjects}
For our investigation, we collaborate with five SMEs. These companies cover various business and technology domains, e.g.~corporate and communal controlling, form letter processing as well as diagnosis and maintenance of embedded systems. Four of them are involved in commercial software development, one in software quality assurance and consulting. 
The latter could not provide an own software project. 

\paragraph{Study Objects (SO)}
Following the suggestion of the partner without a software project, we instead chose the humanitarian open-source system OpenMRS\footnote{\url{http://www.openmrs.org}}, a development of the equally named multi-institution, non-profit collaborative. Hence, our study objects are the five software systems briefly described in Tab.~\ref{fig:studyobjects}. 
These software systems encompass between 100 and 600 kLoC. The developments of SOs 1 to 4 are conducted or audited by the study subjects and started at most seven years ago. The project teams contain less than ten persons. Except for OpenMRS, they are located in the Munich area. The development of SOs 1 and 2 has already been finished before our project started.

\begin{table}[t]
\begin{center}
  \begin{tabular}{c|c|c|c|p{5.1cm}}
    \textbf{SO} & \textbf{Platform} & \textbf{Sources} & \textbf{Size [kLoC]} & \textbf{Business Domain} \\\hline\hline
    1 & C\#.NET & closed, commercial 	& $\approx 100$ & Corporate controlling \\  
    2 & C\#.NET & closed, commercial 	& $\approx 200$ & Embedded device maintenance \\ 
    3 & Java 	& open, non-profit 	& $\approx 200$ & Health information management\\ 
    4 & Java 	& closed, commercial 	& $\approx 100$ & Communal controlling \\ 
    5 & Java 	& closed, commercial 	& $\approx 560$ & Document processing \\ 
  \end{tabular} 
  \caption{Study objects \label{fig:studyobjects}}
\end{center}
\vspace{-10pt}
\end{table}

\subsection{Procedure}
\label{sec:proc}

This section explains milestones of our investigation (Step 1--4). It explains the starting of our research (Step 1), addresses our research questions, i.e.~which data have to be collected and how to achieve that (Step 2) as well as how and under which conditions our analyses have to be carried out (Step 3--4). Steps 2 and 3 take place in terms of a single, collaborative two-week \textit{sprint} per participating enterprise.

\subsubsection{Step 1: Workshops and Interviews}

We conduct a series of workshops and interviews to first convince industrial partners
to participate in our project and then to understand their context and their needs. First, in an information event, we explain the general theme of transferring 
QA techniques and propose first directions.
With the companies that agreed to join the project, we conduct a kick-off meeting
and a workshop to create a common understanding, discuss organisational issues and plan the complete schedule. In addition, the partners
present a software system that we can analyse as well as their needs concerning software quality.
To intensify our knowledge of these systems and problems, for each partner we perform a semi-structured interview with two interviewers and a varying number of interviewees. Both interviewers take notes and consolidate them. We then compare all interview results to find commonalities and differences. Finally, we have one or two consolidation workshops to discuss our results and plan further steps.

\subsubsection{Step 2: Raw Data Collection}
The \textit{source code} of at least three versions of the study objects, e.g.,~major releases chosen by the companies, is retrieved for the application of the chosen techniques for RQ 1 to analyse effects over time. For bug pattern detection and architecture conformance analyses, we retrieve or build executables packed with debug symbols for each of these configurations. For architecture conformance we also need an appropriate \textit{architecture documentation}.
To accomplish this step, all partners have to provide project data as far as available, i.e.~source code, build environment and/or debug builds, as well as documentation of source code, architecture and project management activities.

\subsubsection{Step 3: Measurement and Analysis}
We apply each technique to the gathered raw data via corresponding tool runs and inspect the results, i.e.~findings and statistics. To provide answers for RQ 1 we consider problems arising and efforts spent. The tool runs enable us to derive answers for RQ 2.1. To accomplish this step, the partners have to provide support for technical questions by a responsible contact or by personal attendance at the sprint meetings. One person per technique carries out this step for all SOs. The following explains how this is accomplished:

\paragraph{Code Clone Detection}
\label{sec:proc_clones}

We use the clone detection feature \cite{Juergens2009Clone} of ConQAT 2.7 for all SOs. In case of conventional clone detection the configuration consists of two parameters: the minimal clone length and the source code path. In case of gapped clone detection such gap-specific parameters as maximal allowed number of gaps per clone and maximal relative size of a gap are additionally required. Based on the experience of our group and initial experimentation, we set the minimal clone length to 10 lines of code, the maximal allowed number of gaps per clone to 1 and the maximal relative size of a gap in our analysis to 30\%. After providing the needed parameters we run the analysis.

To inspect the analysis metrics and particular clones we use ConQAT. It provides a list of clones, lists of instances of a clone, a view to compare files containing clone instances and a list of discrepancies for gapped clone analysis. This data is used to recommend corrective actions. Also in a series of runs of clone detection over different versions of respective systems we monitor how several parameters (cf.~Sec.~\ref{sec:methods_clones}) evolve in subsequent versions.

\paragraph{Bug Pattern Detection}
\label{sec:proc_patterns}
For Java-based systems we use FindBugs 1.3.9 and PMD\footnote{\url{http://pmd.sourceforge.net}} 4.2.5. In C\#.NET contexts we use Gendarme\footnote{\url{http://www.mono-project.com/Gendarme}} 2.6.0  and FxCop 10.0. Aside from applying all rules, we choose two additional tool settings which we consider as being relevant for the SOs to gain two focused quality perspectives:
\begin{enumerate}
  \item \textit{Selected categories} addressing correctness, performance, and security
  \item \textit{Selected rules} for unused or poorly partitioned code and bad referencing
\end{enumerate}
The tool settings are determined during preliminary analysis test runs. Categories and rules which are considered as not important -- based on discussion with the partners as well as requirements non-critical to the SOs' application domains -- are ignored during rule selection.

To simplify the issue of defect classification (cf.~Sec.~\ref{sec:methods_patterns}) for our investigation we only distinguish between rules for \textit{bugs} (obvious defects), 
\textit{smells} (simple to very complex heuristics for latent defects) 
and 
\textit{pedantry} (less critical issues with focal point on coding style). 

For additional and language independent metrics (e.g.,~lines of code without comments; code-comment ratio; number of classes, methods and statements; depth of inheritance and nested blocks; comment quality) as well as for result preparation and visualisation we apply ConQAT.

Next, we analyse the finding reports resulting from the tool runs. This step involves the filtering of findings as well as the inspection of source code to confirm the severity and confidence of the findings and to determine corrective actions. To get feedback and to confirm our conclusions from the findings we discuss them with our partners during a workshop.

\paragraph{Architecture Conformance Analysis}
\label{sec:proc_archconf}
 
We use ConQAT for this technique.
The procedure for each system consists of four steps:
\begin{enumerate}
  \item Configuration of the tool with path to source code and corresponding executables of the system
  \item Creation of the architecture reflexion model (cf.~Sec.~\ref{sec:methods_archconf}) based on the architectural information given by the enterprises
  \item Run of the architecture conformance analysis
  \item Defect analysis: Identification, discussion and classification of architectural violations
\end{enumerate}
A detailed description of this ConQAT feature can be found in \cite{Deissenboeck2010}.
In summary, we use a reflexion model where dependency and hierarchy relations between components 
can be expressed. As a next step, we map modelled components to code parts (e.g. packages, namespaces, classes). We exclude code parts from the analysis that do not belong to the system (e.g. external libraries).
Then, ConQAT analyses the conformance of the system to the reflexion model. Every existing dependency that is not allowed by the architectural rules represents a defect. Defects are visualised by the tool on the level of components and on the level of classes and can therefore be analysed on both high and low level.
To eliminate tolerated architecture violations and to validate the created reflexion model, we discuss every 
found defect with the enterprise. As a last step we classify all defects together with the responsible enterprise. 
This allows us to group similar defects and to provide a general understanding.

\subsubsection{Step 4: Questionnaire}
First, we evaluate the experience of the participating enterprises regarding software quality as well as static analysis techniques. Second, we want to understand the perceived usefulness of static analysis techniques for SMEs: Do they plan to use the presented techniques in their future projects? Thus, we perform a survey on our study subjects using a questionnaire containing nine questions (Q1-9), which can be found in Appendix~\ref{sec:questionnairedetails}. This way we contribute to RQ 2.2. The executive managers of each enterprise in their role as a representative for their company then fill out this questionnaire and we evaluate the answers. To avoid the risk of biased or too narrowly formulated answers we use both, open and closed questions.

\section{Results}
\label{sec:results}

We held the information event of Step 1 of our procedure (cf.~Sec.~\ref{sec:proc}) in July 2009 and invited more than thirty SMEs of which finally 12 participated. From these companies, five committed to take part in the project. The other companies were not able to provide the necessary commitment because of schedule or budget constraints. As the first discussions were generally about improving quality assurance, it was not caused by the choice of techniques. We conducted the kick-off meeting in March 2010, the interviews between March and July and, finally, two consolidation workshops in July 2010. We did not particularly analyse their outcome for this paper. But based on these interviews and the experience of our research group, we selected the three static analysis techniques and interpreted our further results.

In the following we portrait for each technique how we contribute to the posed research questions.

\subsection{Code Clone Detection}
\label{sec:results_clones}

\subsubsection{RQ 1.1 -- Technical Problems}
Code clone detection turned out to be the most straightforward and least complicated of the three techniques. It has, however, some technical limitations that could hinder its application in certain software projects.

A major issue was the analysis of projects containing both, markup and procedural code like JSP or ASP.NET. Since ConQAT supports either a programming language or a markup language during a single analysis, it is required to aggregate the results for both languages. To avoid this complication and to concentrate on the code implementing the application logic we took into consideration only the code written in the programming language and ignored the markup code. Nevertheless, it is still possible to combine the results of clone detection of the code written in both languages to get more precise results.

Another technical obstacle was filtering out generated code from the analysed code basis. In one SO large parts of the code were generated by a parser generator, viz.~ANTLR. We excluded this code from our analysis using ConQAT's feature to ignore code files  specified by regular expressions.

\subsubsection{RQ 1.2 -- Spent Effort}

\begin{table}[t]
\begin{center}
  \begin{tabular}{l|L{2.2cm}|L{2.4cm}|L{2.4cm}|L{2.5cm}}
  \textbf{Phase} & \textbf{Work step} & \textbf{(C)lone\newline (D)etection} & \textbf{Bug Pattern Detection} & \textbf{Architecture\newline Conformance} \\\hline\hline
  
  \multirow{3}{2cm}{Introduction (configur\-ation) and calibration}  & Analysis tools                            & $\leq 0.5h$ & $\leq 1h$             & $\leq 0.5h$ \\\cline{2-5}
		                            & Aggregation\newline via ConQAT              & n/a         & $\leq 0.5d$             & $\leq 0.5h$ \\\cline{2-5}
 								 & Recalibration, $x$-times                       & n/a         & $\leq x * 0.5h$             & n/a \\\hline
  \multirow{2}{2cm}{Application (analysis)}       & Run analysis 	                            & $\leq 5min$ & $1 min \leq . \leq 1h$  & $\leq 10sec$ \\\cline{2-5}
		                            & Inspection\newline of results             & $\leq 1h$, more for gapped CD  & $5min \leq . \leq 0.5h$ & $5min \leq . \leq 0.5h$ \\
  \end{tabular} 
  \caption{Efforts spent (RQ 1.2) per study object for applying each of the techniques
  \label{tab:efforts}}
\end{center}
\end{table} 

The effort required to introduce clone detection is small compared to the other two techniques under study. The ease of introduction of clone detection is achieved due to the minimalist configuration of the analysis which in the simplest case includes the path to the source code and the minimal length of a clone. 

For all SOs it took less than an hour to configure clone detection, to get the first results and to investigate the longest and the most frequent clones. Running the analysis process itself took less then five minutes.

In case of gapped clone detection it could take a considerable amount of time to analyse if a discrepancy is intended or if it is a defect. To speed up the process ConQAT supports that the intended discrepancies can be fingerprinted and excluded from further analysis runs. An overview of the efforts can be found in Tab.~\ref{tab:efforts}.

\subsubsection{RQ 2.1 -- Found Defects} 

\begin{table}[t!]
\begin{center}
  \begin{tabular}[\textwidth]{c|c|R{1.6cm}|R{1.3cm}|R{1.5cm}|R{1.5cm}|R{1.5cm}|R{2cm}}

    \textbf{SO} & \textbf{Version} & \textbf{Analysed Units [kUnits]} & \textbf{Cloned Units [kUnits]} &\textbf{Blow-up\newline [\%]} & \textbf{Unit\newline Coverage [\%]} & \textbf{Longest Clone [Units]} & \textbf{Most Clone Instances} \\\noalign{\hrule height 1pt}
      & I   &  15,9 &  3,5 & 119.5 & 22.2 & 112 &  39 \\\cline{2-8}  
    1 & II  &  25,3 &  5,8 & 118.9 & 23.0 & 117 &  39 \\\cline{2-8}    
      & III &  32,3 &  7,8 & 119.2 & 24.0 & 117 &  39 \\\noalign{\hrule height 1pt}
      & I   &  35,4 & 14,3 & 143.1 & 40.5 &  63 &  64 \\\cline{2-8}  
    2 & II  &  41,6 & 18,9 & 150.2 & 45.4 & 132 &  47 \\\cline{2-8}    
      & III &  39,9 & 14,6 & 137.4 & 36.7 &  89 &  44 \\\noalign{\hrule height 1pt}
      & I   &  51,7 &  9,4 & 114.5 & 18.2 &  79 &  21 \\\cline{2-8}  
    3 & II  &  56,8 &  8,6 & 111.2 & 15.1 &  52 &  20 \\\cline{2-8}    
      & III &  61,6 &  8,4 & 110.0 & 13.7 &  52 &  19 \\\noalign{\hrule height 1pt}
      & I   &   8,9 &  6,0 & 238.8 & 68.0 & 217 &  22 \\\cline{2-8}  
    4 & II  &  22,4 & 17,3 & 309.6 & 77.6 & 438 &  61 \\\cline{2-8}    
      & III &  38,3 & 30,4 & 336.0 & 79.4 & 957 & 183 \\\noalign{\hrule height 1pt}
      & I   & 196,3 & 48,7 & 122.3 & 24.8 & 141 &  72 \\\cline{2-8}  
    5 & II  & 211,3 & 53,4 & 122.7 & 25.3 & 158 &  72 \\\cline{2-8}    
      & III & 208,6 & 53,2 & 122.8 & 25.5 & 156 &  72 \\\noalign{\hrule height 1pt}
  \end{tabular} 
  \caption{Results of code clone detection
   \label{fig:ucstudyobjects}}
\end{center}
\vspace{-10pt}
\end{table}

The results of conventional clone detection can be interpreted as an indicator of bad design or of bad software maintainability, but they do not point at actual defects. Nevertheless, these results give first hints, which code parts must be improved. The following three design flaws were detected in all analysed systems to a certain extent: cloning of exception handling code, cloning of logging code and cloning of interface implementation by different classes.

Tab.~\ref{fig:ucstudyobjects} shows the clone detection results for three versions of each SO, sorted by time. In the analysed systems unit coverage as defined in Sec.~\ref{sec:methods} varied between 14 and 79\%. Koschke \cite{Koschke2007} reports on several case studies with unit coverage values between 7 and 23\% and one case study with a value of 59\%, which he defines as extreme. Therefor, the SOs 1, 3 and 5 contain normal clone rates according to Koschke. The clone rate in SO 2 is higher than the rates reported by Koschke and for SO 4 it is extreme. Regarding maintenance the calculated blow-up for each system is an interesting value. For example version III of SO 4 is more than three times bigger as its hypothetically equivalent system containing no clones. SO 4 shows that cloning can be an increasing factor over time, while SO 3 reveals that it is possible to reduce the amount of clones existing in the system code. 

\begin{table}[t!]
\begin{center}
  \begin{tabular}[\textwidth]{c|c|R{1.6cm}|R{1.3cm}|R{1.5cm}|R{1.5cm}|R{1.5cm}|R{2cm}}
    \textbf{SO} & \textbf{Version} & \textbf{Analysed Units [kUnits]} & \textbf{Cloned Units [kUnits]} &\textbf{Blow-up\newline [\%]} & \textbf{Unit\newline Coverage [\%]} & \textbf{Longest Clone [Units]} & \textbf{Most Clone Instances} \\\noalign{\hrule height 1pt}
      & I   &  13,3 &  3,0 & 119.9 & 22.3 &  34 &  39 \\\cline{2-8}  
    1 & II  &  21,0 &  4,5 & 117.9 & 21.5 &  37 &  52 \\\cline{2-8}    
      & III &  27,1 &  6,0 & 117.4 & 22.1 &  52 &  52 \\\noalign{\hrule height 1pt}
      & I   &  24,3 &  4,6 & 116.3 & 19.0 & 156 &  37 \\\cline{2-8}  
    2 & II  &  34,7 &  8,7 & 123.2 & 25.0 & 156 &  37 \\\cline{2-8}    
      & III &  37,1 &  9,4 & 123.7 & 25.3 & 156 &  37 \\\noalign{\hrule height 1pt}
      & I   &  46,7 & 12,0 & 124.4 & 18.2 &  73 & 123 \\\cline{2-8}  
    3 & II  &  46,1 & 10,0 & 120.0 & 15.1 &  55 &  67 \\\cline{2-8}    
      & III &  49,1 & 10,0 & 118.6 & 20.5 &  55 &  64 \\\noalign{\hrule height 1pt}
      & I   &   7,8 &  4,5 & 192.1 & 58.6 &  42 &  34 \\\cline{2-8}  
    4 & II  &  18,8 & 11,0 & 206.2 & 59.8 &  51 &  70 \\\cline{2-8}    
      & III &  32,2 & 19,2 & 211.1 & 59.5 &  80 & 183 \\\noalign{\hrule height 1pt}
      & I   & 142,3 & 29,4 & 117.4 & 20.7 &  66 &  68 \\\cline{2-8}  
    5 & II  & 154,0 & 32,8 & 118.0 & 21.3 &  85 &  78 \\\cline{2-8}    
      & III & 151,9 & 32,7 & 118.2 & 21.5 &  85 &  70 \\\noalign{\hrule height 1pt}
  \end{tabular} 
  \caption{Results of gapped code clone detection
  \label{fig:gappedclones}}
\end{center}
\vspace{-10pt}
\end{table} 

Cloning is considered harmful because it increases the chance of unconscious, inconsistent changes, which can lead to faults in a system~\cite{Juergens2009}. These changes can be detected when applying gapped clone detection. We found a number of such changes in the cloned code fragments, but we could not classify them as defects, because we lacked the knowledge needed about the software systems. Also the project partners could not directly classify these discrepancies as defects, which confirms that gapped clone detection is a more resource demanding type of analysis. Nevertheless, in some clone instances we identified additional instructions or deviating conditional statements compared to other instances of the same clone class. Gapped clone detection does not go beyond method boundaries, since experiments showed that inconsistent clones that cross method boundaries in many cases did not capture semantically meaningful concepts~\cite{Juergens2009}. This explains why metrics such as cloned units or clone coverage may differ from values observed with conventional clone detection. Tab.~\ref{fig:gappedclones} shows the results of gapped clone detection.

\subsubsection{RQ 2.2 -- Perceived Usefulness} 

Following the feedback obtained from the questionnaire, two enterprises had limited experience with clone detection, the others did not know about it at all (Q2). Three enterprises estimated the relevance of clone detection to their projects as very high, the others estimated it as medium relevant (Q3). Concerning Q3, one stated that ``clones are necessary within short periods of development.'' Finally, all enterprises evaluated the importance of using clone detection in their projects as medium to high and plan to introduce this technique in the future (Q5).
For details see Tables \ref{tab:rq2.2_1} and \ref{tab:rq2.2_2} in Appendix \ref{sec:questionnairedetails}.

\subsection{Bug Pattern Detection}
\label{sec:results_patterns}

\subsubsection{RQ 1.1 -- Technical Problems}
Following Sec.~\ref{sec:methods_patterns}, we confirm that bug patterns are a powerful technique to gather a vast variety of information about potentially defective code. However, most of its effectiveness and efficiency is achieved through carefully done, project-specific fine-tuning of the many setscrews available.

First, the impact of findings on quality factors of interest and their consequences for the project (e.g.~corrective actions, avoidance or tolerance) were difficult to determine by the tool-provided rule categories, the severity and confidence information. Based on our experience we identified the following study object characteristics this impact depends on:
\begin{itemize}
  \item Required usage-level qualities, e.g., security, safety, performance, usability 
  \item Required internal qualities, e.g., code maintainability, reusability 
  \item Technologies, i.e.,~language, framework, platform, architectural style
  \item Criticality of the context the findings belong to, e.g., platform or driver code 
\end{itemize}

Second, some rules exhibited many false positives, either because their technical way of detection is fuzzy or because a definitely precise finding is considered not relevant in a project-specific context. The latter case requires an in-depth understanding of each of the rules, the impacts of findings and, subsequently, a proper redlining of rules as pedantry or, actually, irrelevant. We neither measured the rates of false positives nor investigated costs and benefits thereof as our focus lay on the identification of the most important findings only. 

Third, due to restricted selection and filtering mechanisms in the tools as well as a bounded view of the SOs' life-cycles, we were hindered to apply and calibrate appropriate rule selectors and findings filters. We saw that the usefulness of results is crucially influenced by the conversion of project-specific information on rule impacts into queries for rule selection and findings filtering.
The tools greatly differ in their abilities to achieve this task via their graphical or command-line interfaces. 

We addressed the first two issues by group discussion also with our partners and improved rule selection and findings filtering to principally avoid the findings reports to get overloaded or prone to false positives of the second kind. Also, the third issue could only be largely compensated by manual efforts. As most finding reports were quite homogeneously encoded and technically well accessible, we utilised ConQAT to gain statistical information for higher-level quality metrics as listed in Step 3 of Sec.~\ref{sec:proc_patterns}.

\subsubsection{RQ 1.2 -- Spent Effort}

We achieved the initial setup of a single bug pattern tool in less than an hour. This step required knowledge about the internal structure of the SO such as, e.g.,~its directory structure and third party code.
We used the ConQAT framework to flexibly run the tools in a specific setting (Java only) and for further processing of the finding reports. Having good knowledge of this framework, we completed the analysis setup for an SO (selection of rules, adjustment of bug pattern parameters, framework setup, etc.) in about half a day.

The runs took between a minute and an hour depending on code size, rules selection and other parameters. Hence, bug pattern detection should at least be selectively included into automated build tasks. Part of the rules are computationally complex and some tools frequently required more than a gigabyte of memory. 
The manual effort after the runs can be split into review and recalibration.
The review of a report took us a few minutes up to half an hour.
Due to the short period of the life-cycle of the SOs we had insight into, we could not estimate the recalibration effort for the rule selector and the findings filter. An overview of the efforts can be found in Tab.~\ref{tab:efforts}.

\subsubsection{RQ 2.1 -- Found Defects}
We conducted bug pattern analysis in three selective tool settings according to Step 3 in Sec.~\ref{sec:proc_patterns}, but only for one version of each SO. For all SOs the filtered finding reports confirmed the defects focused or expected by these settings. Without going into the quantities and details of single findings, we summarise language-specific results:
\begin{description}
  \item[\textbf{C\#}] Upon the rules with highest numbers of findings, FxCop and Gendarme reported \textit{empty exception handlers}, \textit{visible constants}, and \textit{poorly structured code}. There was only one consensually critical kind of findings related to correctness in SO 2, viz.~unacceptable loss of precision through \emph{wrong cast during an integer division} used for accounting calculations.

  \item[\textbf{Java}] Upon the rules with highest numbers of findings, FindBugs and PMD reported \textit{unused local variables}, \textit{missing validation of return values}, \textit{wrong use of serialisable}, and extensive \textit{cyclomatic complexity}, \textit{class/method size}, \textit{nested block depth}, \textit{parameter list}. There have only been two consensually critical findings, both in SO 5, related to correctness, viz.~foreseeable \emph{access of a null pointer} and an \emph{integer shift beyond 32 bits} in a basic date/time component.
\end{description}
Independent of the programming language and concerning security and  stability 
we frequently detected the pattern \textit{constructor calls an overwritable method} in 4 of 5 SOs and found a number of defects related to \textit{error prone handling of pointers}.
Concerning maintainability the SOs exhibited \textit{missing or unspecific handling of exceptions}, manifold \textit{violation of code complexity metrics} and various forms of \textit{unused code}. Details are shown in Tab.~\ref{fig:bugpatternstats}.

\begin{table}[t!]
\begin{center}
  \begin{tabular}{p{1.3cm}|p{5cm}|c|c|c|c|c|p{2.4cm}}

\multirow{2}{1.2cm}{\textbf{Tool (Lang.)}} & \multirow{2}{4.9cm}{\textbf{Rule}\newline(recommendations in parentheses)} & \multicolumn{5}{c|}{\textbf{Study Objects}} & \multirow{2}{2.5cm}{\textbf{Most affected Qualities}} \\
& & 
$\;\;1\;\;$ &    
$\;\;2\;\;$ &    
$\;\;3\;\;$ &    
$\;\;4\;\;$ &    
$\;\;5\;\;$ &    
\\\hline\hline
\multirow{2}{1.3cm}{FxCop (C\#)} & Empty / general exception handlers  					
	& 47 	 & 106 & n & n & n & Maintainability \\
& Nested use of generic types   				
	& 44  &  .  & n & n & n & Maintainability \\\hline
\multirow{2}{1.3cm}{Gend\-arme (C\#)} & Deep namespaces				
	& 35  &	.  & n & n & n & Maintainability \\
	    & Visible constants  			
	& 18  & 338 & n & n & n & Security \\
 		& Extensively large classes    	
	& . &	3 & n  & n  & n  & Maintainability \\
          & Extensively long methods 
	& .  &  17 & n  &	n & n & Maintainability \\
     	& Suspicious type conversion   	  	
	& .  &   3 &	n &	n & n & Correctness \\\hline
Gend., PMD & Constructor calls overwritable method 
	&  8  &	. & x &  x  & x  & Security, stability \\\hline
\multirow{2}{1.3cm}{Find\-Bugs (Java)} & Unused local variables				
	&  n  &	n & 142 &  .  &  .  & Maintainability \\
 	& Inefficient string manipulation
	&  n  &	n & 46  &  .  & .  & Performance \\
 		 & Corrupted serialisable		
	&  n  &	n & 55  &  .  & .  & Correctness \\
		& Return values not validated		
	&  n &  n  & 30  &  .  & .  & Correctness, sec.\\ 
     & Access of a null pointer		
	&  n  & n  & .   &  .  &  1  & Sec., stability \\
     & Integer shift beyond 32 bits 			
	&  n  &  n  &  .  &  .  &  4  & Correctness \\\hline
\multirow{2}{1.3cm}{PMD (Java)}  & Empty method in abstract class	
	&  n  &  n  &  x & . &  x & Maintainability \\
        & Max.~cyclomatic complexity ($\le 10$)
	&  n  &  n  &  78 & 156 
	  			   	  & 216 & Maintainability \\
        & Extensive length / size / parameter count, too many methods / fields	   
	&  n	 &  n  &  .  & x & x & Maintainability \\\hline
 ConQAT & Max.~nested block depth ($\le 5$)		
	&  13 & 11 &  19 & 17 
				   	   & 14 & Maintainability \\
  \end{tabular} 
  \caption{Overview of bug pattern results. Legend: Cells contain the number of findings or a maximum value, ``n'' \dots not applicable, ``.'' \dots not noticeable, ``x'' \dots noticeable, but PMD did not offer an appropriate way to exactly count the many findings \label{fig:bugpatternstats}}
\end{center}
\vspace{-10pt}
\end{table}

\subsubsection{RQ 2.2 -- Perceived Usefulness}

According to the questionnaire, all of the partners considered our bug pattern findings to be medium to highly relevant for their projects (Q3).
The sample findings we presented during our final workshop were perceived as being non-critical for the success of the SOs but would have been treated if they had been found by such tools during the development of these software systems.
The low number of consensually critical findings correlated well with the fact that the technique was known to all partners and that most of them have good knowledge thereof and regularly used such tools in their projects, i.e.~at least monthly, at milestone or release dates (Q1-2). 
However, three of them could gain additional education in this technique (Q4).
Nevertheless, all of the enterprises decided to use bug patterns as an important QA instrument in their future projects (Q5). 
For details see Tab.~\ref{tab:rq2.2_1} and \ref{tab:rq2.2_2} in Appendix \ref{sec:questionnairedetails}.

\subsection{Architecture Conformance Analysis}
\label{sec:results_archconf}

\subsubsection{RQ 1.1 -- Technical Problems}
We observe two kinds of general problems that prevent or complicate each architectural analysis: The absence of an architecture documentation and the usage of dynamic patterns.

For two of the systems there was no documented architecture available. In one case the information was missing because the project was taken over from a different organisation that was not documenting the architecture at all. They reasoned that any later documentation of the system architecture would be too expensive for their enterprise. In another case the organisation was aware that their system was severely lacking any architectural documentation. Nevertheless they feared that the time involved and the sheer volume of code to be covered exceeds the benefits. The organisation additionally argued that they are afraid of having to update the documentation within several months as soon as the next release is coming out.

In SO 2 a dynamic architectural pattern is applied, 
where nearly no static dependencies could be found between defined components. All components belonging to the system are connected at run-time. Thus, our static analysis approach could not be applied.

Architecture conformance analysis needs two ingredients apart from the architecture documentation: The source code and the executables of a system. This could be a problem because the source has to be compilable to analyse it.
Another technical problem occurred when using ConQAT. Dependencies to components solely existing as executables were not recognised by the tool. For that reason all rules belonging to compiled components could not be analysed. 

Beside these problems we could apply our static analysis approach to two systems without any technical problems. An overview of all SOs with respect to their architectural properties can be found in Tab.~\ref{fig:archstudyobjects}.

\subsubsection{RQ 1.2 -- Spent Effort}
For each system the initial configuration of ConQAT and the creation of the reflexion model in ConQAT could be done in less than one hour. Tab.~\ref{fig:archstudyobjects} shows the number of modelled components and the rules that were needed to describe their allowed connections. The analysis process itself finished in less than ten seconds. The time needed for the interpretation of the analysis results is of course dependent on the amount of defects found. For each defect we were able to find the causal code parts within one minute. We expect that the effort needed  for bigger systems will only increase linearly but staying small in comparison to the benefit that can be achieved using architecture conformance analysis as illustrated in Sec.~\ref{sec:methods_archconf}. An overview of the efforts can be found in Tab.~\ref{tab:efforts}.

\begin{table}[t!]
\begin{center}
  \begin{tabular}{c|l|c|C{4.2cm}|C{3.2cm}}
    \textbf{SO} & \textbf{Architecture} & \textbf{Version} & \textbf{Violating Component Relationships} & \textbf{Violating Class Relationships}  \\ \hline\hline
      & \multirow{3}{0.2\textwidth}{12 Components\newline 20 Rules} & I & 1 & 5 \\\cline{3-5}
    1 &  & II & 3 & 9 \\\cline{3-5}
      & & III & 2 & 8 \\\hline
    2 & dynamic & n/a & n/a & n/a \\\hline
    3 & undocumented & n/a & n/a & n/a \\\hline
      &  \multirow{3}{0.2\textwidth}{14 Components 9 Rules}& I & 0 & 0 \\\cline{3-5}
    4 & & II & 1 & 1 \\\cline{3-5}
      & & III & 2 & 4 \\\hline
    5 & undocumented & n/a & n/a & n/a \\\hline
  \end{tabular} 
  \caption{Architectural characteristics of the study objects \label{fig:archstudyobjects}}
\end{center}
\vspace{-10pt}
\end{table} 

\subsubsection{RQ 2.1 -- Found Defects}

As shown in Tab.~\ref{fig:archstudyobjects} we observed several discrepancies in the analysed SOs over nearly all version. Only one version did not contain architectural violations. 
Overall, we found three types of defects in the analysed systems. Each defect represents a code location showing a discrepancy to the documented architecture. The two analysable SOs had architectural defects which could be avoided if this technique had been applied. 
In the following we explain the types of defects we classified together with the responsible enterprises. 
The companies rated all findings as critical.
 
\begin{itemize}
\item  \textit{Circumvention of abstraction layers:} Abstraction layers (e.g. presentation layer) provide a common way to structure a system into logical parts. The defined layers are hierarchically dependent on each other, reducing the complexity in each layer and allowing to benefit from structural properties like exchangeability or flexible deployment of each layer. These benefits vanish when the layer concept is harmed by dependencies between layers that are not connected to each other. In our case e.g. the usage of the data layer from the presentation layer was a typical defect we found in the analysed systems. 

\item \textit{Circular dependencies:} We found undocumented circular dependencies between two components. We consider these dependencies -- whether or not documented -- as defects themselves, because they affect the general principle of component design. Two components that are dependent on each other can only be used together and can thus be considered as one component, which contradicts the goal of a well designed architecture. The reuse of these components is strongly restricted. They are harder to understand and to maintain. 

\item \textit{Undocumented use of common functionality:} Every system has a set of common functionality (e.g. date manipulation) which is often grouped into components and used across the whole system. Consequently, it is important to know where this functionality is actually used inside a system. Our observation showed that there were such dependencies that were not covered by the architecture.

\end{itemize}

\subsubsection{RQ 2.2 -- Perceived Usefulness}
Following the feedback gained from the questionnaire, we observed that 4  of the 5 participating enterprises did not know about the possibility of automated architecture conformance analysis (Q1). Only one of them already checked the architecture of their systems, however in a manual way and less frequently. Confronted with the results of the analysis all enterprises rated the relevance of the presented technique medium to highly relevant (Q3). One of them stated that as a new project member it is easier to become acquainted with a software system if its architecture conforms to its documented specification. All enterprises agreed on the usefulness of this technique and plan its future application in their projects (Q5).
Details of the questionnaire can be found in Tab.~\ref{tab:rq2.2_1} and \ref{tab:rq2.2_2} in Appendix \ref{sec:questionnairedetails}.

\section{Discussion}
\label{sec:discussion}

\paragraph{General Observations}

First, we observed that code clone detection and architecture conformance analysis have been quite new to our partners as opposed to bug pattern detection which was well known.
This may result from the fact that style checking and simple bug pattern detection are  standard features of modern development environments.
However, we consider it as important to know that code clone detection can indicate critical and complex relationships residing in the code at minimum effort.
We made our partners aware of the usefulness of architecture conformance analysis, both in the case of an available architecture specification and to reconstruct such a documentation.

Second, we conclude that all of the three techniques can be introduced and applied with resources affordable for small enterprises.
We assume, that except for calibration phases at project initiation or after substantial product changes the effort of readjusting the settings for the techniques stays very low. This effort is compensated by the time earned through narrowing results to successively more relevant findings.
Moreover, our partners perceived all of the discussed techniques as useful for their future projects.

Third, we perceive our analyses of the study objects as successful. We found large clone classes, a significant number of pattern-based bugs aside from smells and pedantry as well as unacceptable architecture violations.

\paragraph{Usage Guidelines}
During the repetitive conduct of Steps 2 and 3 of the procedure in Sec.~\ref{sec:proc} we gained a lot of experience in applying the chosen techniques. For their introduction and application to a new software project we consider the following generic procedure as very helpful:
\begin{enumerate}
  \item Establish a project-specific \textit{configuration}. This includes the choice, particularly for bug patterns, of appropriate rules aiming on relevant quality factors or just the strengthening of design or coding guidelines.

  \item Define 
    events for \textit{measurement, findings filtering and documentation}. Filtering requires in-depth knowledge of the system and its critical components. For bug pattern detection this influences severity and confidence levels, and for architecture conformance analysis this influences the definition of allowed, tolerated, and forbidden dependencies.
  
  \item Decide whether to \textit{treat or tolerate} findings. This involves (i) the inspection of results and defective code, (ii) the issue of change requests for defect removal and, (iii) to assess efficiency, the documentation of efforts spent.

  \item Determine whether and how defects can be \textit{avoided} regarding lessons learned from defect treatment.

  \item Strengthen \textit{quality gates} by improved criteria, which follow patterns such as, e.g.,\
    ``Clone coverage in critical code package $A$ below $X\%$ prior to any bundled feature introduction.'', 
    ``No critical security errors with confidence $> Y\%$ according to tool $Z$ for any release.'', or
    ``No architecture violations originating from change sets of new features.''

  \item For \textit{project control} in the context of \textit{continuous integration}, derive statistics and trends from findings reports by a quality control dashboard such as ConQAT.
\end{enumerate}

\section{Threats to Validity}
\label{sec:threats}
 
In the following, we discuss threats to the validity of our results. 
We structure them in internal and external validity
threats.
 
\subsection{Internal Validity} 

First, a potential threat to the internal validity is that most of the project
participants had little experience with the specific tools we were applying.
This could give us additional technical problems, which would not have
occurred with experts. Furthermore, the efforts are probably higher. We
mitigated this risk by discussions with experts and we assume that the
introduction in other companies would also not necessarily be performed by 
experts.

Second, we did not record exact details about the efforts we spent.
We rather made order of magnitude estimations only. In our context we consider this threat as small as we do not require precise analyses of these efforts including time measurement.

Third, we did not completely check whether all defects we found have caused real problems such as, e.g., critical system failures during operation or significant budget overruns. Hence, there may be false positives. We reduced this risk by detailed inspections of the defects we listed.

Fourth, the questionnaire results could be wrong, because a participant either knowingly or unknowingly gave incorrect answers. We mitigated this threat by asking participants to be careful in filling it out and at the same time assured anonymity to them.

\subsection{External Validity} 

As this is an experience report on a technology transfer project, the results are inherently
difficult to generalise. We had five projects of SMEs all located in Germany.
We also restricted our analysis to systems realised in Java and C\# and only applied specific analysis tools for it.
Hence, the problems, defects, and perceptions may be particular to this context.

Nevertheless, we think that most of our experiences are valid for other contexts as
well. The companies, we have collaborated with, range in their size from only several to
a hundred employees. The domains they build software for differ quite strongly. Finally,
the tools are all prominent examples and had been used in industrial projects before.
Only the restriction to two programming languages has a strong effect as for other languages
there may exist rather different tools and defects. For instance, with bug pattern detection,
Ahsan, Ferzund and Wotawa~\cite{DBLP:conf/icsea/AhsanFW09} report that characteristics 
of bug patterns may be language specific.

\section{Related Work}
\label{sec:relatedwork}

In this research we concentrate on applying automated static analysis techniques to enable SMEs mitigate the risk of defect-related costs. Different from our approach, the research community devotes its attention primarily to software process improvement in SMEs. There are a number of papers covering this topic. 

Kautz \cite{Kautz1999} developed and used metrics to evaluate how new practices and tools for configuration and change management were affecting the software process at three SMEs. This work considers that the key to successful software measurement is to make metrics meaningful and to tailor them to a particular organisation. We confirm that observation in the context of software measurement.

Von Wangenheim et al. \cite{Wangenheim2006} investigated the assessment of software processes in SMEs to improve these processes. They developed MARES, a set of guidelines for conducting an ISO/IEC 15504-conforming software process assessment, focused on small companies. 
We perceive the usage guidelines we reported 
as a potential bridge between automated static analysis and more general guidelines for software process improvement.

Hofer \cite{hofer2002} states that only 10\% of the analysed SMEs in Austrian software industry believe to suffer from a lack of methods. He concludes that appropriate tool support as well as the knowledge of methods is available. On the contrary, we argue that SMEs may not be aware of many effective methods and can therefore not estimate their lack concerning these techniques.

Returning to automated static analysis techniques, to the best of our knowledge, multiple techniques have never been applied in a study in an SME context. However, there are several publications in which such techniques were investigated separately and in other contexts:

Lague et al. \cite{Lague1997} report on application of function clone detection to a large telecommunication software system. As opposed to that, we do not limit clone detection to the comparison of functions but compare arbitrary code fragments with each other. In this work we also did not analyse large systems. Nevertheless, we came to the similar conclusion that clone detection has potential to improve software quality.

Lanubile and Mallardo \cite{Lanubile2003} performed research on finding clones in web applications developed using markup and programming languages. As mentioned earlier, 
our approach is technically limited in analysing such software systems. Introducing a semi-automatic approach presented by Lanubile and Mallardo could remove this limitation. 

Ayewah et al.~\cite{ayewah07} evaluate the accuracy and value of FindBugs findings and discuss but not solve the problem of properly filtering false positives. They use the term \emph{trivial bugs} 
for what we call smells and pedantry. We confirm their conclusions on the usefulness of findings and believe that an application of bug pattern detection has to undergo calibration guided by the staff of a software project.
Moreover, by answering RQ2, we contribute to Foster's, Hicks' and Pugh's \cite{foster2007improving} question ``Are the defects reported by [static analysis] tools important?''.

Ferzund, Ahsan and Wotawa \cite{DBLP:conf/iwsm/FerzundAW08} report on the effectiveness of rules for smell detection. The rules they developed are based on machine learning 
and source file statistics provided by static code metrics. They used training information from two software projects including bug databases. We did not address the estimation of rule effectiveness but focused on their selection and application.

Wagner et al.~\cite{DBLP:conf/icst/WagnerDAWS08} similarly applied FindBugs and PMD to two industrial projects. They could not find defects reported from the field that are covered by bug pattern detection. However, our results show that this technique indeed captures critical defects that may eventually occur in the field.
 
Rosik et al.~\cite{Rosik2008industrial} conducted an industrial case study on architecture conformance with three participating software engineers. They conclude that this technique should be integrated into the software engineering process and applied continuously. We think that the procedure we presented is able to satisfy their needs, because it explicitly focuses on continuous integration.

Mattsson et al.~\cite{Mattsson2007} illustrate their experience in an industrial project and the huge effort that is needed to keep the architectural model in conformance with the implementation. However, they tried to reach this goal in a manual way. Our results show that automation can dramatically reduce efforts.

Feilkas, Ratiu and Juergens~\cite{Feilkas2009loss} analysed three .NET platform projects of Munich Re very similar to our procedure, but they analysed the effects of the loss of architectural knowledge. Compared to our results they report a much higher effort of about five days to apply the technique, mainly because of time consuming discussions. We think that the lower effort we are reporting is mainly caused by the fact that we were collaborating with small enterprises and experienced a lower communication overhead.

\section{Conclusions and Future Work}
\label{sec:conclusions}

In general, it is most effective to combine different QA
techniques to find most of the defects~\cite{Littlewood:2000:MEC:358134.357482}. This, however, comes at the
efforts and costs of performing many different techniques. Particularly, SMEs have difficulties in assigning large
efforts to diverse QA provisions and to training specialists for them. Automated
static analysis techniques promise to be an efficient contribution to software QA, because they only require little effort for their application.

We reported our experience in applying three static analysis techniques to small enterprises: code clone detection, bug pattern detection and architecture conformance analysis. Consequently, we assessed potential barriers for introducing these techniques as well as the observations we could make in a one-year project with five German SMEs.

We found several technical problems, such as multi-language projects with single
language clone analysis or false positives, but we believe that these are no major
road blocks for the adoption of static analysis. Overall, the effort for introducing the
analyses was small. Most techniques were set up with an effort of less than one person-hour.
We found various defects, such as high levels of cloning, null pointer access, erroneous calculations or circumvention of architecture layers. In the end, our partners found all of the presented techniques relevant for inclusion into their quality assurance processes.

In our opinion static analysis tools can efficiently improve quality assurance in SMEs, 
if they are continuously used throughout the development process and are technically well 
integrated into the tool landscape. But as our research was not focused on long term observations 
we can not address this issue. Consequently it is an interesting area of future work to investigate 
the long term effects of static analyses in SMEs' software projects and their continuous integration into their development 
processes. Questions arising from the application of these techniques such as their long term efficiency, their inclusion into an overall QA strategy, their acceptance by developers, their application to non-code development artefacts or their effects
on the daily work could then be investigated.

We will continue to work in this area to better understand the needs of SMEs and investigate our current findings.

\vspace{-5pt}

\section*{Acknowledgements}

We would like to thank Christian Pfaller, Bernhard Sch\"atz and Elmar J\"urgens for their technical and organisational support throughout the project. 
We thank all involved companies for their reproach-less collaboration and assistance.
\vspace{-5pt}

\bibliographystyle{abbrv}

\pagebreak
\begin{appendix}
\appendix
\label{sec:appendix}

\section{Results of the Questionnaire}
\label{sec:questionnairedetails}

\begin{table}[h!]
\begin{center}
  \footnotesize
  \begin{tabular}{p{4.2cm}|l}
  \textbf{Question} & \textbf{Closed Answers} (without comments) \\\hline\hline
  
  \textbf{Q1)} Which of these static analysis tech\-niques have you already been using in your projects? &
  \begin{tabular}[t]{l|p{.5cm}p{.5cm}p{.5cm}p{.5cm}r}
    & \rb{daily} & \rb{weekly} & \rb{monthly} & \rb{less freq.} & \rb{never} \\
    Architecture conformance & 0 & 0 & 0 & 1 & 4 \\
    Bug pattern detection    & 2 & 2 & 1 & 0 & 0 \\ 
    Clone detection          & 0 & 0 & 0 & 2 & 3 
  \end{tabular}
  \\\hline
  
  \textbf{Q2)} What is your estimate of the experience of your company in these techniques? &
    \begin{tabular}[t]{l|p{.5cm}p{.5cm}p{.5cm}p{.5cm}r|c}
	    & ++ & + & o & -- & - - & none \\
	    Architecture conformance & 1 & 2 & 1 & 1 & 0 & 0 \\
	    Bug pattern detection    & 1 & 3 & 1 & 0 & 0 & 0 \\
	    Clone detection          & 0 & 0 & 1 & 0 & 1 & 3 
    \end{tabular}
  \\\hline
  
  \textbf{Q3)} How do you perceive the relevance of our analysis results for your study object? & 
	\begin{tabular}[t]{l|p{.5cm}p{.5cm}p{.5cm}p{.5cm}r|c}
	    & \multicolumn{2}{l}{high} & o & \multicolumn{2}{r|}{low} & none \\
	    Architecture conformance & 3 & & 2 & & 0 & 0 \\
	    Bug pattern detection    & 2 & & 3 & & 0 & 0 \\
	    Clone detection          & 3 & & 2 & & 0 & 0 
	\end{tabular}
  \\\hline
  
  \textbf{Q4)} How much education could you gain from the topics of our research project? & 
	\begin{tabular}[t]{l|p{.5cm}p{.5cm}p{.5cm}p{.5cm}r|c}
	    & \multicolumn{2}{l}{much} & o & \multicolumn{2}{r|}{little} & none \\
	    Architecture conformance & 2 & 2 & 1 & 0 & 0 & 0 \\
	    Bug pattern detection    & 2 & 0 & 1 & 1 & 1 & 0 \\
	    Clone detection          & 2 & 2 & 1 & 0 & 0 & 0
	\end{tabular}
  \\\hline

  \textbf{Q5)} Which of the following analysis techniques do you plan to apply at which level of priority? &
	\begin{tabular}[t]{l|p{.5cm}p{.5cm}p{.5cm}p{.5cm}r|c|c}
	    & ++ & + & o & -- & - - & none & *) \\
	    Architecture conformance & 1 & 3 & 0 & 1 & 0 & 0 & 5 \\
	    Bug pattern detection    & 4 & 1 & 0 & 0 & 0 & 0 & 5 \\
	    Clone detection          & 0 & 2 & 3 & 0 & 0 & 0 & 5 
	\end{tabular} \\
	& *) application of the technique is planned
  \\\hline
 
  \end{tabular} 
  \caption{Summary of closed answers of the questionnaire for RQ 2.2 (five results, contents and answers have been translated from German to English). Legend: ++ .. very high, + .. high, o .. medium, -- .. low, - - .. very low
  \label{tab:rq2.2_1}}
\end{center}
\vspace{-10pt}
\end{table}

\begin{table}[t!]
\begin{center}
  \footnotesize
  \begin{tabular}{p{.99\textwidth}}
  \textbf{Open Answers and Comments} \\\hline
  
  \textbf{Q1)}
  Architecture conformance analysis has not been used because \dots
  \begin{itemize}
    \item ``projects have been developed cleanly or without [need of] architecture.''
    \item ``manual inspection was carried through.''
    \item ``the prerequisites \dots would have needed to be established for our projects. Manual inspection (code reviews) already takes place irregularly.''
    \item ``it was not known to us.''
  \end{itemize}

  Clone detection has not been used because \dots
  \begin{itemize}
    \item ``[clones were] not known to us as a problem.''
    \item ``we did not recognise its necessity.''
  \end{itemize}
  \\
  
  \textbf{Q3)}
  The results have been relevant because \dots
  \begin{itemize}
    \item ``manual [code] analysis is significantly more cost-intensive, \dots clone detection is only feasible with tool support.''
    \item ``we learned about concepts, experiences and tools \dots it is easier to become acquainted with [a project if its architecture conforms to its documented specification].''
  \item ``Clones are necessary within short periods of development.''
  \end{itemize}
  \\
  
  \textbf{Q5)}
	``The results of this research project shall be included into our internal development process.''
  \\

  \textbf{Q6)} \textit{Your estimate of the current status of your organisation w.r.t.~software quality:} 
  
  \textit{Strengths:} ``Seamless process for requirements QA \dots regarded design guidelines for all languages used \dots flexible adaptation of guidelines to customer needs \dots performed QA provisions (from unit testing to selective pair programming) seem to work \dots
  so far we only experienced high customer satisfaction \dots mature in testing techniques and management.''
  
  \textit{Weaknesses:} ``No consequent QA provisions \dots no systematic QA \dots automation and tool usage either project specific or even left out \dots still learning to apply the tools.''
  \\

  \textbf{Q7)} \textit{Where do you expect the highest potential of your organisation to improve its software quality?} 
  \begin{itemize}
    \item ``Consequent QA provisions,''
    \item ``integrated tools and more automation \dots QA dashboard for project managers,''
    \item ``better knowledge transfer between teams and projects,''
    \item ``improved quality control \dots backflow of QA results into development process.''
  \end{itemize}
  \\

  \textbf{Q8)} \textit{Your estimate of the usefulness of static analysis for your software projects:}
  
  \textit{Positive:} ``Important'', ``high'', ``trend analyses are important'', ``very important, because of early and efficient defect detection \dots help identify structural deficits \dots ease [code] maintenance \dots quality improvement starting with first build \dots for internal projects better control and indication of deficits.''

  \textit{Negative:} ``Often not feasible in projects externally conducted at the customers'.''
  \\\hline
  
  \end{tabular} 
  \caption{Summary of open answers and comments of the questionnaire for RQ 2.2 (five results, contents and answers have been translated from German to English)
  \label{tab:rq2.2_2}}
\end{center}
\vspace{-10pt}
\end{table}

\end{appendix}

\end{document}